\documentclass[aps,onecolumn,preprint]{revtex4}

\usepackage{graphicx}
\usepackage{dcolumn}
\usepackage{natbib}
\usepackage{algorithm}
\usepackage{algorithmic}
\usepackage{amssymb}

\begin{document}

\title{A Fast and Efficient Algorithm for Slater Determinant Updates in Quantum Monte Carlo Simulations}

\author{Phani K.V.V. Nukala}
\affiliation{Computer Science and Mathematics Division, 
Oak Ridge National Laboratory, Oak Ridge, TN 37831-6164, USA}
\author{P. R. C. Kent}
\affiliation{Center for Nanophase Materials Sciences, 
Oak Ridge National Laboratory, Oak Ridge, TN 37831-6164, USA}

\begin{abstract}
  We present an efficient low-rank updating algorithm for updating the
  trial wavefunctions used in Quantum Monte Carlo (QMC)
  simulations. The algorithm is based on low-rank updating of the
  Slater determinants. In particular, the computational complexity of
  the algorithm is $\mathcal{O}(k N)$ during the $k$-th step compared
  with traditional algorithms that require $\mathcal{O}(N^2)$
  computations, where $N$ is the system size. For single determinant
  trial wavefunctions the new algorithm is faster than the traditional
  $\mathcal{O}(N^2)$ Sherman-Morrison algorithm for up to
  $\mathcal{O}(N)$ updates. For multideterminant
  configuration-interaction type trial wavefunctions of
  $M+1$ determinants, the new algorithm is significantly
  more efficient, saving both $\mathcal{O}(MN^2)$ work and
  $\mathcal{O}(MN^2)$ storage. The algorithm enables more accurate and
  significantly more efficient QMC calculations using configuration
  interaction type wavefunctions.
\end{abstract}

\maketitle

\section{Introduction}\label{sec:intro}

Quantum Monte Carlo (QMC) is an approach capable of yielding highly
accurate results in systems ranging from isolated molecules to the solid
state\cite{WMCFoulkesRMP2001}. 
The success of most common QMC methods, namely variational 
Monte Carlo (VMC) and diffusion Monte Carlo (DMC), depends crucially on the choice of
trial wavefunction. Indeed, the trial wavefunction limits both the
statistical efficiency and accuracy of the simulation. In QMC methods, 
the evaluation of the trial wavefunction becomes the most
demanding part of the calculation especially when sufficiently large systems are 
considered or accurate simulations are required. This aspect of QMC was recognized even in the earliest DMC calculations, e.g. Ref. \cite{PJReynoldsJCP1982}. Consequently, the choice of trial
wavefunction used in QMC calculations is motivated both by the
accuracy and the speed of evaluation.

The most common form of trial wavefunction is of the Slater-Jastrow type
\begin{equation}
  \label{eq:1}
  \Psi({\bf R})=D({\bf R})e^{J({\bf R})},
\end{equation}
where, neglecting spin, $D({\bf R})$ is a Slater determinant, 
$J({\bf R})$ is a Jastrow function, and
${\bf R}=\left\{{\bf r}_1,{\bf r}_2, ..., {\bf r}_N\right\}$ is a
vector of the position ${\bf r}_i$ of each electron. 
In QMC, the simulation commonly proceeds by proposing a local change to the
electronic system configuration ${\bf R}$ to ${{\bf R}^\prime}$. This local
change in ${\bf R} \mapsto {\bf R}^\prime$ is induced by
the movement of one electron at a time from position ${\bf r}_i$ to ${\bf r}_i^\prime$.
The probability that the proposed local change
is accepted is dependent on the transition probability, which
depends on the ratio of $\Psi({\bf R}^\prime)/\Psi({\bf R})$, where ${\bf R}^\prime$
is a new set of electron positions. This transition probability computation in turn 
requires the computation of the ratio of 
determinants  ${ D}({\bf R}^\prime)$ and
${ D}({\bf R})$ in the new and old configurations respectively.
Although a complete re-computation of $D({\bf R}^\prime)$ can be made, 
an efficient algorithm that computes the 
necessary ratio $D({\bf R}^\prime)/D({\bf R})$ without resorting to a 
complete independent calculation of each determinant can significantly 
increase the overall efficiency of QMC simulations. Indeed, 
this efficiency measure is essential to the success
of Slater-Jastrow wavefunctions; for a single electron move, the conventional 
algorithms use Sherman-Morrison formula 
(special case of Sherman-Morrison-Woodbury formula \cite{GHGolubBookMatrixComputations1996}) 
which reduces the cost of evaluating $D({\bf R}^\prime)/D({\bf R})$ to $\mathcal{O}(N)$, with an
$\mathcal{O}(N^2)$ cost if the move is accepted, compared with
$\mathcal{O}(N^3)$ for a naive evaluation of the determinant. Once the ratio of the determinants has been calculated, most quantities required in the Monte Carlo can be obtained through a simple multiplicative scaling\cite{WMCFoulkesRMP2001}. Comparatively recently, ``linear scaling'' approaches have been developed to reduce the cost of evaluating the determinants\cite{AJWilliamsonPRL2001,FAReboredoPRB2005,DAlfeJPC2004,AAspuruGuzikJCC2005,JKussmanPRB2007} by exploiting spatial locality in the studied physical system. In this paper, we explore alternative and complementary approaches to speedup the computation of transition probabilities and determinant ratios in QMC calculations.

The most accurate and commonly used QMC method is the DMC method performed in the fixed node approximation.
This method exhibits a varational error in the energy depending on the quality of the nodal 
surfaces (zeroes) of the trail wavefunction. To improve the nodes as well as the
variational quality of the trial function, it is now routine to
utilize multiple determinant trial functions. These are commonly
obtained from multiconfiguration quantum chemistry approaches such as
the configuration interaction method where the ground state
determinant $D$ is supplemented by single and double
excitations from the ground state. That is, 

\begin{equation}
  \label{eqn:mdwf} \Psi({\bf R})=\left[D({\bf R})+\sum_{a,c}
\alpha_{a,c}D_{a}^{c}({\bf R})+\sum_{a,b,c,d} \beta_{a,b,c,d}D_{ab}^{cd}({\bf
R})+...\right]e^{J({\bf R})},
\end{equation}
where $D_{ab}^{cd}$ denotes a double excitation with
orbitals $a$ and $b$ replaced by $c$ and $d$ respectively, and $\alpha$ and $\beta$ 
denote the multi-determinant expansion coefficients. 
Higher order excitations may be progressively included. Such an expansion of the 
wavefunction allows the nodal surface to be improved.

There are many strong motivations for minimizing the computational cost of
multideterminant wavefunctions in QMC:
Recent benchmark tests of the accuracy achievable in all electron VMC
utilized, for example, up to 499 determinants to obtain over 90\% of
the correlation energy in the first row atoms\cite{MDBrownJCP2007}. To
obtain a similar fraction of correlation energy in larger systems,
more determinants are likely required. Numerous recent studies 
\cite{JWLawsonCPL2008,JToulouseJCP2008,JRTrailJCP2008} have
shown the utility of increased numbers of determinants for improved
accuracy in atomic, molecular, and solid-state
applications. In general this result is expected since quantum
chemical techniques systematically improve the wavefunction with
increased numbers of determinants. Improved trial wavefunctions using
multideterminants are required for large systems such as the $C_{60}$
fullerene where current trial wavefunctions are insufficient for computing 
accurate optical properties\cite{MLTiagoJCP2008}. Multiple
determinants may also be required to represent certain spin
symmetries, e.g. Ref. \cite{RQHoodPRL2003}. Additionally, we have also
recently shown that it is possible to sample the ground state
wavefunction into a configuration expansion\cite{FAReboredoPRB2008}
and subsequently improve the trial
wavefunction\cite{FAReboredoPRB2009Unpub}. This application requires
the use of large configuration interaction expansions consisting of
potentially thousands of determinants.

In this paper we propose an efficient algorithm for utilizing
Slater-Jastrow trial wavefunctions in QMC simulations. The algorithm
is particularly efficient for multideterminant
wavefunctions. Extension to related alternative wavefunction forms
such as multi-pfaffian and multi-backflow wavefunctions is
straightforward. In Section \ref{sec:alg} we present the details of
the algorithm. Section \ref{sec:bench} presents benchmark timing and
efficiency measures for single determinant calculations using a
variety of system sizes.  The multideterminant case is analysed in
Section \ref{sec:mdetana}. Conclusions are given in Section
\ref{sec:conclusions}.

\section{Algorithms for updating Slater determinants}
\label{sec:alg}
As mentioned earlier, in QMC, 
the Monte Carlo simulation proceeds by proposing a local change to the 
electronic system configuration ${\bf R}$ to ${{\bf R}^\prime}$. 
The acceptance criterion for each such local change follows the 
traditional Metropolis algorithm, which requires the computation 
of the transition probability. Each time a local 
change is accepted, the Slater matrix ${\bf D}({\bf R})$ is updated  
to ${\bf D}({\bf R}^\prime)$ by modifying one of the rows of ${\bf D}({\bf R})$ 
corresponding to an electron movement from ${\bf r}$ to ${\bf r}^\prime$.
The simulation then proceeds by proposing 
a new local change, which requires the re-computation of the determinant 
of Slater matrix ${\bf D}({\bf R}^{\prime\prime})$ in the subsequent 
configuration ${\bf R}^{\prime\prime}$. This progression of the simulation 
via local changes typically proceeds for many thousands to  millions of steps 
until observables such as the total energy converge to 
a desired statistical accuracy.

The slater matrix in configuration ${\bf R} = ({\bf r}_1, {\bf r}_2, \cdots, {\bf r}_N)$ 
is given by 
\begin{eqnarray}
{\bf D}({\bf R}) & = & \left[ \begin{array}{cccc}
\phi_1({\bf r}_1) & \phi_2({\bf r}_1) & \cdots & \phi_N({\bf r}_1) \\
\phi_1({\bf r}_2) & \phi_2({\bf r}_2) & \cdots & \phi_N({\bf r}_2) \\
\vdots & \vdots & \ddots & \vdots \\
\phi_1({\bf r}_N) & \phi_2({\bf r}_N) & \cdots & \phi_N({\bf r}_N) \\
\end{array} \right] \label{Apsi}
\end{eqnarray}
where ${\bf r}_i$ and $\phi_i$ for $i = 1,2,\ldots,N$ indicate respectively the spatial coordinates and 
spin-orbitals of $i$-th electron. Because we are moving a single electron (say $p$-th electron) 
at a time from position ${\bf r}_p \mapsto {\bf r}^\prime_p$, the Slater matrix in the 
new electronic configuration ${{\bf R}^\prime} = ({\bf r}_1, {\bf r}_2, \cdots, {\bf r}^\prime_p, \cdots, {\bf r}_N)$ 
is simply obtained by modifying the $p$-th row as 
\begin{eqnarray}
{\bf D}({\bf R}^\prime) & = & \left[ \begin{array}{cccccc}
\phi_1({\bf r}_1) & \phi_2({\bf r}_1) & \cdots & \phi_p({\bf r}_1) & \cdots & \phi_N({\bf r}_1) \\
\phi_1({\bf r}_2) & \phi_2({\bf r}_2) & \cdots & \phi_p({\bf r}_2) & \cdots & \phi_N({\bf r}_2) \\
\vdots & \vdots & \ddots & \vdots & \vdots & \vdots \\
\phi_1({\bf r}^\prime_p) & \phi_2({\bf r}^\prime_p) & \cdots & \phi_p({\bf r}^\prime_p) & \cdots & \phi_p({\bf r}^\prime_p) \\
\vdots & \vdots & \vdots & \vdots & \ddots & \vdots \\
\phi_1({\bf r}_N) & \phi_2({\bf r}_N) & \cdots & \phi_p({\bf r}_N) & \cdots & \phi_N({\bf r}_N) \\
\end{array} \right] \label{Apsip}
\end{eqnarray}
The Metropolis probability to accept or reject the move is dependent on the ratio of determinants 
of Slater matrices $R = \frac{D({\bf R}^\prime)}{D({\bf R})}$, 
where $D({\bf R}^\prime)$ is the determinant of Slater matrix ${\bf D}({\bf R}^\prime)$ and 
$D({\bf R})$ is the determinant of ${\bf D}({\bf R})$. 
The transition probability is in general proportional to $|R|^2$, assuming real wavefunctions. If the move 
is accepted, then the system configuration changes to ${\bf R}^\prime$; if not, the move is 
rejected and the system remains in configuration ${\bf R}$. 
In the following, whenever the 
context is clear, we denote $D({\bf R}^\prime)$ by $D^\prime$ and $D({\bf R})$ by D.

For the Monte Carlo simulation to be efficient, all quantities related
to the transition probability and any observables must be computed
with minimum computational operations. In the case of a single
determinant wavefunction the ratio $D^\prime/D$ is required. For a
single electron move this corresponds to a change of a single
row in the Slater matrix. However, for the case of the multideterminant wavefunction, as in
Eq. \ref{eqn:mdwf}, all ratios $D_{a}^{c\prime}/D_{a}^{c}$ and
$D_{ab}^{cd^\prime}/D_{ab}^{cd}$ are required. These ratios
involve determinants with both orbital replacements and single electron moves 
(i.e., both row and column changes) when compared to the original ground state
determinant $D$.
 
For a single electron move, 
the basic computational problem involved during the $(k+1)$-th MC step may be expressed as: 
\emph{Given the determinanat $D_k$ of Slater matrix ${\bf D}_k$, compute the determinant $D_{k+1}$ 
of ${\bf D}_{k+1}$} such that 
\begin{eqnarray}
{\bf D}_{k+1} & = & {\bf D}_k + {\bf e}_{{\bf p}(k)} {\bf v}_{k}^t \label{eq1}
\end{eqnarray}
where ${\bf p}(k)$ defines an index vector that maps $k \mapsto p$ such that 
${\bf p}(k) = p$, and ${\bf e}_{p}$ denotes an unit vector with $1$ on the 
$p$-th entry and $0$ everywhere else. The vector ${\bf v}_k$ corresponds to the 
{\it change} in Slater matrix due to the 
displacement of the $p$-th electron during the $(k+1)$-th MC step and is 
given by 
\begin{eqnarray}
{\bf v}_k & = & \left(\phi_1({\bf r}^\prime_p)-\phi_1({\bf r}_p),  
\cdots, (\phi_N({\bf r}^\prime_p)-\phi_N({\bf r}_p)\right)^t \label{vk}
\end{eqnarray}
A straight-forward computation of $D_{k+1}$ may be obtained as
\begin{eqnarray}
D_{k+1} & = & (1 + {\bf v}_{k}^t {\bf D}_{k}^{-1} {\bf e}_{{\bf p}(k)})~D_{k} \label{Aeq}
\end{eqnarray}
and $R_k = \frac{D_{k+1}}{D_k}$ can be evaluated as 
\begin{eqnarray} 
R_k & = & (1 + {\bf v}_{k}^t {\bf D}_{k}^{-1} {\bf e}_{{\bf p}(k)}) \label{Req}
\end{eqnarray}

Hence, for any given $k$, $R_k$ can be evaluated efficiently in $\mathcal{O}(N)$ computations since 
${\bf D}_{k}^{-1} {\bf e}_{{\bf p}(k)}$ can be interpreted as the $p$-th column of 
${\bf D}_{k}^{-1}$, i.e., ${\bf D}_{k}^{-1} (:,p) = {\bf D}_{k}^{-1} {\bf e}_{{\bf p}(k)}$. 
However, repetitive computation of $R_k$ during 
each of the MC simulation steps (for $k = 0,1,2,\ldots$) requires an efficient procedure to compute 
${\bf D}_{k}^{-1}$ for each $k$. 
For this purpose, traditional algorithms employ the Sherman-Morrison formula to update 
${\bf D}_{k}^{-1} \mapsto {\bf D}_{k+1}^{-1}$, which can be expressed as 
\begin{eqnarray}
{\bf D}_{k+1}^{-1} & = & {\bf D}_{k}^{-1} - \frac{{\bf D}_{k}^{-1} {\bf e}_{{\bf p}(k)} {\bf v}_k^t {\bf D}_{k}^{-1}}
{(1 + {\bf v}_k^t {\bf D}_{k}^{-1} {\bf e}_{{\bf p}(k)})} \label{rank1G}
\end{eqnarray}
Using this formula, ${\bf D}_{k}^{-1}$ can be updated to ${\bf
  D}_{k+1}^{-1}$ in $\mathcal{O}(N^2)$ computations. 
However, since the required number of MC
steps in a typical Monte Carlo simulation readily extends to the
thousands to millions range, and can increase with increasing system
sizes, $\mathcal{O}(N^2)$ scaling of these traditional algorithms
poses a significant hindrance for the simulation of large system sizes
despite the fact that such large scale simulations are necessary to
develop a better understanding of relevant chemistry and physics. As discussed in
the introduction, Sec \ref{sec:intro}, multiple determinants compound
this problem.

Alternatively, an efficient recursive algorithm for computing ${\bf D}_{k+1}^{-1}$ 
may be formulated by expressing Eq. \ref{rank1G} as
\begin{eqnarray}
{\bf D}_{k+1}^{-1} & = & \left[{\bf I} - \frac{{\bf D}_{k}^{-1} {\bf e}_{{\bf p}(k)} {\bf v}_k^t}
{(1 + {\bf v}_k^t {\bf D}_{k}^{-1} {\bf e}_{{\bf p}(k)})} \right] ~ {\bf D}_{k}^{-1} \nonumber \\
& = & \left[{\bf I} - \gamma_k {\bf u}_k {\bf v}_k^t \right] ~ {\bf D}_{k}^{-1}
\label{rank1Ga}
\end{eqnarray}
where ${\bf u}_k = {\bf D}_k^{-1} {\bf e}_{{\bf p}(k)}$ and 
$\gamma_k = \frac{1}{R_k}$. Based on Eq. \ref{rank1Ga}, a recursive 
scheme for computing ${\bf D}_{k+1}^{-1}$ may be formulated as 
\begin{eqnarray}
{\bf D}_{k+1}^{-1} & = & ({\bf I} - \gamma_k {\bf u}_k {\bf v}_{k}^t) \ldots 
({\bf I} - \gamma_0 {\bf u}_0 {\bf v}_{0}^t) ~ {\bf D}_{0}^{-1} \nonumber \\
& = & \left[\prod_{j=0}^k ({\bf I} - \gamma_j {\bf u}_j {\bf v}_{j}^t) \right]~ {\bf D}_{0}^{-1} 
\label{rank1Gb}
\end{eqnarray}
An $\mathcal{O}(k N)$ recursive algorithm based on Eq. \ref{rank1Gb}
is presented in Algorithm \ref{alg:QMC}. For each additional step $k$,
this algorithm requires storage space for two vectors ${\bf u}_k$ and
${\bf v}_k$ of size $N$. In addition, we need to store an index vector
${\bf p}(k)$ that maps $k \mapsto p$ such that ${\bf p}(k) = p$.

\section{Single determinant benchmarks}
\label{sec:bench}

In order to compare the computational efficiency of the recursive algorithm
with the traditional algorithm, we first tested  the case of a single determinant wavefunction. 
An analysis of the multideterminant case is given in Sec. \ref{sec:mdetana}.

\begin{figure}
  \centering
\includegraphics{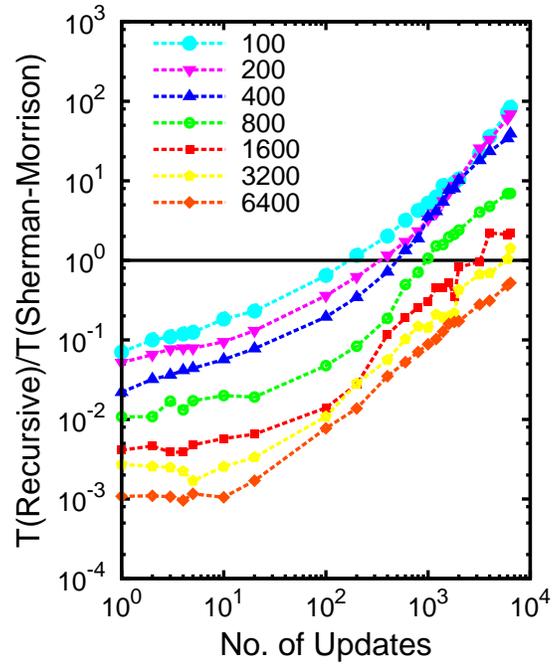}  
  \caption{Relative timing of the recursive update algorithm to the traditional Sherman-Morrison algorithm for different matrix sizes. Ratios less than one indicate that the recursive algorithm is faster.}
  \label{fig:timings}
\end{figure}

We tested the algorithms on a randomly generated matrix ${\bf D}_0$. 
That is, since the algorithms are applicable for general
matrices, we start with a matrix ${\bf D}_0$ whose elements are
randomly chosen between zero and one. Then we consider rank-1 updates
of ${\bf D}_0$ as given by Eq. \ref{eq1} for $m$ number of
steps. The site locations $p$ are chosen sequentially, modulo $N$, for
these $m$ steps. The updated orbitals are chosen randomly.  Figure
\ref{fig:timings} presents the ratio of the computational timings
obtained using the full matrix updating and recursive updating
algorithms. The timings were obtained using a standalone benchmark
code using double precision arithmetic. We used the same data
structures both in our recursive and full QMC simulations. Machine optimized linear
algebra library calls were used for both algorithms.
Timings were obtained on a 2.73 GHz Intel Xeon processor with 12 MB Cache. 

Examining the timings shown in Fig. \ref{fig:timings}, we see that the
recursive update algorithm is always significantly faster than the
Sherman-Morrison algorithm for a small number of updates. For up to
ten updates, the new algorithm $\sim 10$ times faster for a 100 sized
matrix, while for a 6400 sized matrix the new algorithm is $\sim1000$
times faster. For increased numbers of updates the ratio of timings
decreases. The crossover between the two algorithms occurs near the
theoretically expected $k=N$ updates.

\begin{figure}
  \centering
\includegraphics{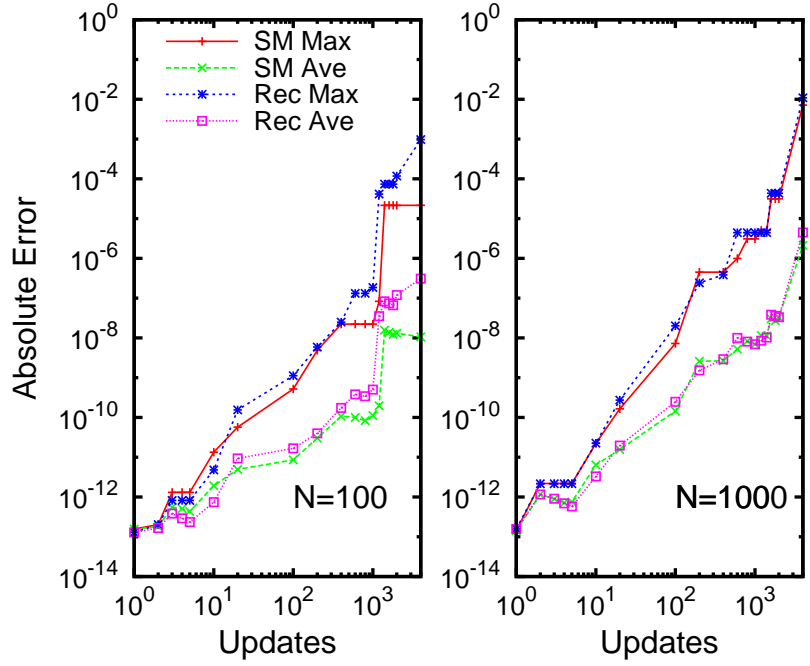}  
  \caption{Absolute numerical errors in computed determinant ratios using the recursive update (Rec) and Sherman-Morrison (SM) algorithms with double precision arithmetic for matrix sizes of 100 (left) and 1000 (right). The behavior of the algorithms is similar.}
  \label{fig:errors}
\end{figure}

Due to the iterative nature of both algorithms numerical errors
accumulate over time. It is common practise in QMC simulations to
fully recalculate the inverse cofactor matrices from time to time to
limit these errors. Such a recalculation requires $\mathcal{O}(N^3)$ operations. We have compared
the numerical errors of the recursive update algorithm with the
Sherman-Morrison algorithm and find the performance to be
similar. Figure \ref{fig:errors} illustrates the build up of errors
for both algorithms for a single run.

Figure \ref{fig:errors} shows that both algorithms have good stability
and on average give high accuracy, particularly for small numbers of
updates. However, for both algorithms the average and maximum
numerical error in the determinant ratio gradually increases with the
number of updates and can become substantial. In both cases the
maximum error for a fixed number of updates can deviate by several
orders of magnitude from the average. This behavior appears to be due
to the occasional mixture of very small and very large numbers in the
update formulae which results in a significant loss of precision. This
data shows that while the recursive algorithm performs similarly to
the Sherman-Morrison algorithm, it is vital to check sufficient
accuracy is obtained if large numbers of updates are performed.

\section{Multiple determinant wavefunctions}
\label{sec:mdetana}

In the case of multiple determinant wavefunctions such as a
configuration interaction expansion, all the excited Slater matrices ${\bf D}_a^c({\bf R})$ 
and ${\bf D}_{ab}^{cd}({\bf R})$ are similar
to the ground state matrix ${\bf D}({\bf R})$, and differ only by a few column
interchanges. The use of the recursive algorithm provides an efficient 
way of calculating the transition probability compared to the traditional
algorithm; It is not only faster but also requires reduced storage of $\mathcal{O}(N^2)$ for 
storing \emph{only ${\bf D}_0^{-1}$ of the ground state matrix}. No other
potentially large data must be stored, although it is advantageous to
reuse the current determinant values between MC steps. The recursive
algorithm is used to compute the non-ground state determinants via column
changes to the ground state matrix. The cost of each particle move is
constant and does not increase when many steps are taken.

For simplicity we analyse the case of a multiple determinant
wavefunction consisting of only the ground state determinant and $M$
determinants doubly excited from this state. Conventionally the
${\bf D}_0^{-1}$ as well as all the excited Slater matrix inverses are 
stored in memory to enable fastest possible update using the traditional algorithm. When the 
recursive algorithm is applied to multiple determinant wavefunctions, we 
store only the ${\bf D}_0^{-1}$ of the ground state. Conventional updates are 
performed on this determinant and the recursive algorithm is used to compute
the other excited determinants since the excited and ground state 
Slater matrices differ by a few column changes. 
Note that successive row updates can be performed in $\mathcal{O}(k N)$ 
operations using an algorithm similar to that of Algorithm \ref{alg:QMC}. However, 
successive row updates followed by multiple column updates always requires 
a $\mathcal{O}(N^2)$ cost associated with a matrix-vector multiplication. 
Since proposed moves are usually accepted
in DMC calculations with an acceptance ratio of $> 99 \%$, it is convenient 
to use the conventional (Sherman-Morrison) algorithm to update the inverse of 
the ground state Slater matrix. 
It should also be noted that in the event the proposed move is accepted, 
the traditionally updated $D({\bf R}^\prime)$ used in the evaluation of 
the excited state determinants can be reused: the
recursive update algorithm then requires no additional
$\mathcal{O}(N^2)$ work over a single determinant calculation. 
Consequently, using the recursive algorithm an $M$
determinant wavefunction can be used with an updating cost scaling
only linearly in $M$ and system size $N$ compared to an $N^2$ 
scaling cost using the traditional algorithm.

To evaluate determinant ratios such as $D_{ab}^{cd}({\bf R^\prime})/D_{ab}^{cd}({\bf R})$ 
we first perform a traditional update to obtain $D({\bf R^\prime})$. 
The recursive algorithm is then used to compute
$D_{ab}^{cd}(\bf R^\prime)$ from $D(\bf R^\prime)$. We assume that
$D_{ab}^{cd}(\bf R)$ is stored and available from a previous MC step, but this
can also be calculated using two applications of the recursive
algorithm to $D(\bf R)$. 
In Table \ref{tab:mdcosts} we compare the costs of evaluating the
determinant ratios in $\Psi({\bf R^\prime})/\Psi({\bf
  R})$. Independent of the amount of storage chosen for the
traditional scheme, the recursive scheme displays an improved computational cost by
a factor $\mathcal{O}(MN^2)$, or $M$ times the cost of a complete
single determinant update. The single determinant benchmarks of
Sec. \ref{sec:bench} show that these updates, which are few in number and hence 
correspond to the left side of Fig. \ref{fig:timings}, are several orders of
magnitude faster than the traditional algorithm.

\section{Conclusions}
\label{sec:conclusions}
In this paper, we presented an efficient low-rank updating algorithm
for QMC simulations.  The algorithm requires only $\mathcal{O}(k N)$
computations during $k$-th MC step compared with $\mathcal{O}(N^2)$
computations required by traditional algorithms. Our numerical
simulations indicate that for small numbers of updates of a single
determinant this algorithm is orders of magnitude faster than
traditional algorithms. For single determinant wavefunctions, the
traditional algorithms remain the preferred choice for more than
$\mathcal{O}(N)$ updates. For multideterminant wavefunctions of $M+1$
determinants, our algorithm is the preferred choice, being
significantly faster and of particular interest for large systems. In addition, it 
enables workspace to be reduced by a factor $\mathcal{O}(MN^2)$. The
speed and storage savings of this new algorithm enables QMC
calculations to use thousands of determinants.

\par
\vskip 1.00em%
\noindent {\bf Acknowledgment} \\ PRCK wishes to thank F. A. Reboredo,
J. Kim, and R. Q. Hood for helpful conversations. This research is
sponsored by the Mathematical, Information and Computational Sciences
Division, Office of Advanced Scientific Computing Research and the Center for Nanophase Materials Sciences, Office of Basic Energy Sciences, both of the
U.S. Department of Energy and under contract number DE-AC05-00OR22725 with
UT-Battelle, LLC. The QMC Endstation project is supported by the
U.S. Department of Energy (DOE) under contract number DOE-DE-FG05-08OR23336.

\newpage
\bibliographystyle{apsrev}

\newpage

\begin{algorithm}
\caption{Recursive Algorithm ($p$-th electron moves)}
\begin{algorithmic}[1]
\STATE Given ${\bf D}_0^{-1}$ and ${\bf v}_k$
\STATE Set ${\bf p}(k) = p$ 
\STATE Set ${\bf u}_k = {\bf D}_0^{-1} {\bf e}_{{\bf p}(k)} = {\bf D}_0^{-1}(:,p)$
\FOR{$i = 0$ to $k-1$}
	\STATE Compute ${\bf u}_{k} = {\bf u}_{k} - \gamma_i~({\bf v}_i^t {\bf u}_k)~{\bf u}_i$
\ENDFOR
\STATE Compute $R_k = 1 + {\bf v}_k^t {\bf u}_k$
\IF {Accept}
\STATE Compute $\gamma_k = \frac{1}{R_k}$
\STATE Save ${\bf u}_k$, ${\bf v}_k$ and $\gamma_k$
\STATE k = k+1
\ENDIF
\end{algorithmic}
\label{alg:QMC}
\end{algorithm}

\newpage

\newpage

\newpage
\begin{table}
\begin{ruledtabular}
  \begin{tabular}{cccc}
    Algorithm & Move evaluation cost & Move acceptance cost & Storage cost    \\ \hline
Traditional                    & $\mathcal{O}((1+M) N)$ & $\mathcal{O}((1+M) N^2)$  & $\mathcal{O}(2(1+M)N^2)$ \\
Minimum storage traditional    & $\mathcal{O}((1+2M)N^2+(1+M)N)$& $\mathcal{O}(N^2)$        &$\mathcal{O}(2N^2)$      \\
Recursive                      & $\mathcal{O}(N^2+3MN)$ & $\mathcal{O}(N^2)$        & $\mathcal{O}(2N^2)$      \\ \hline
  \end{tabular}
\end{ruledtabular}
\caption{Cost of computing wavefunction ratios using traditional and recursive algorithms for 
proposed and accepted single electron moves. The wavefunction is of the configuration interaction 
doubles type consisting of $M$ double excitations from a single ground state determinant. For the 
storage costs we consider only the most significant $\mathcal{O}(N^2)$ and higher contributions. 
For at least the ground state determinant, both the full matrix and its inverse are stored 
resulting in the lead factor of 2 in the storage costs. In the traditional algorithm, 
the emphasis is on speed and hence all the excited state matrices and 
ground state matrix (along with its inverses) are stored. 
For the ``minimum storage traditional'', the emphasis is on limiting storage costs even 
at the expense of increased computational cost. Hence, in the "minimum storage traditional" 
algorithm, we assume that the traditional algorithm is used but 
only the ground state matrix is stored and the remaining matrices are computed based on the 
ground state matrix.}
  \label{tab:mdcosts}
\end{table}
\end{document}